\documentclass{appolb}
\usepackage{epsfig}

\begin{document}
\title{Recent developments in modeling neutrino interactions in 1~GeV energy region%
\thanks{Presented at the Cracow Epiphany Conference {\it On Physics in Underground Laboratories and its Connection with LHC}, Cracow, January 6-8, 2010}%
}
\author{Jan T. Sobczyk
\address{Institute of Theoretical Physics\\ Wroc\l aw University}
}
\maketitle
\begin{abstract}
Recent experimental and theoretical research in the area of neutrino interactions in the $\sim 1$~GeV region are reviewed including topics like: the problem of value of quasielastic axial mass, neutral current $\pi^0$ production, coherent pion production. Many comments are devoted to status and current development of Monte Carlo events generators.

\end{abstract}

\section{Introduction}

Knowledge of neutrino interactions in the $\sim~1$~GeV energy region is important because this energy domain is typical for majority of neutrino oscillations experiments performed during recent $\sim 5$ years and also those scheduled for the near future. The list includes K2K, MiniBooNE, SciBooNE, MINOS, T2K and NO$\nu$A. The only exception is OPERA with higher energy neutrino beam from CERN.\\
\\
Neutrino oscillations are the energy dependent phenomenon and the most straigthforward analysis of experimental data requires reconstruction of neutrino energy. The neutrino flux spectrum is typically rather wide-band (despite significant improvements introduced with the idea of off-axis beams) and the interacting neutrino energy must be estimated based on the observation of the leptonic and/or hadronic final states. The precision of the  analysis depends on the knowledge of neutrino interaction cross-sections, both inclusive and  exclusive in several most important channels. If the analysis of the oscillation signal does not include as the intermediate step the neutrino energy reconstruction \cite{sobczyk_zmuda}, it is still based on the comparison with predictions from the Monte Carlo (MC) events generators and rely on how well they are known and implemented in numerical codes. In the analysis of the oscillation appearance signal as seen in Cherenkov detectors it is very important to evaluate the background from NC$\pi^0$ production events. Since the required precision of new oscillation experiments is higher with respect to what was sufficient until recently the series of dedicated workshops was initiated with the aim to organize a forum of discussion for experimentalists and theorists, including specialists in nuclear physics, to present new measurements and proposed improvements in models \cite{nuint}. Thanks to NuInt workshops during the last 8 years the knowledge about neutrino interactions improved significantly  even if some problems and limitations turned out to be quite robust. \\
\\
In the $\sim 1$~GeV energy region one distinguishes several processes that invoke quite different theoretical descriptions. The terminology is sometimes confusing because neutrino reactions can take place both on free nucleons and  (this is the common situation) nucleus targets. Fortunately in the $\sim 1$~GeV energy region typical values of momentum transfer are large enough and with a good approximation one can assume that neutrino-nucleus reaction occurs on individual bound nucleons (impulse approximation - IA). Limitations of this picture will be addressed several times in what follows. Thus we distinguish: quasielastic (QE) CC reaction 
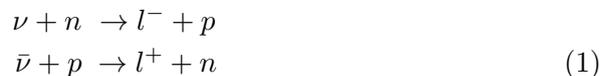
\begin{eqnarray}
\nu + n &\rightarrow l^- + p\nonumber \\
\bar \nu + p &\rightarrow l^+ + n
\end{eqnarray}
and its NC elastic counterpart

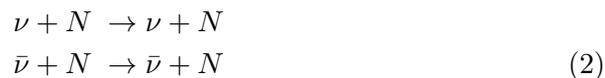
\begin{eqnarray}
\nu + N &\rightarrow \nu+ N\nonumber\\
\bar\nu + N &\rightarrow \bar\nu+ N
\end{eqnarray}
single-pion production (SPP) reactions

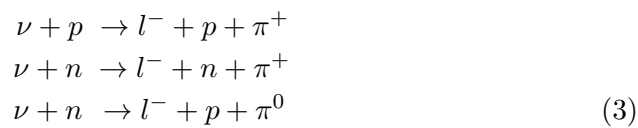
\begin{eqnarray}
\nu + p &\rightarrow l^- + p + \pi^+\nonumber\\
\nu + n &\rightarrow l^- + n + \pi^+\nonumber\\
\nu + n &\rightarrow l^- + p + \pi^0
\end{eqnarray}
(with many other channels for anti-neutrinos and for NC reactions) and more inelastic reactions (the misleading name {\it DIS} is commonly used). The seperation of three types of reaction is quite general and the specific feature of the $\sim 1$~GeV region is that all three (QE, SPP, DIS) contributions to the total cross-section are important. The necessity to have a consistent inclusion of both SPP and DIS dynamics is a source of many difficulties in the construction of MC generators of neutrino interactions. In the case of neutrino-nucleus reaction there is  a possibility also of a coherent pion production (COH)

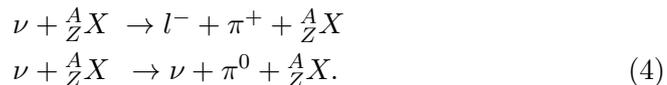
\begin{eqnarray}\label{coh}
\nu +{}^A_ZX &\rightarrow l^- + \pi^+ +{}^{A}_{Z}X\nonumber\\
\nu +{}^A_ZX &\rightarrow \nu + \pi^0 +{}^{A}_{Z}X.
\end{eqnarray}
The second reaction contributes to the overall NC $\pi^0$ production and is a subject of intensive experimental and theoretical investigation.\\
\\
Nuclear effects make this picture more complicated. The first obvious and nontrivial difficulty is the problem of nuclear enviroment modifications of free neutrino-nucleon interactions. The second problem comes from final state interactions (FSI). Even if we accept as reasonable the assumption that the reaction takes place on individual nucleons, the particles that arise in the final state propagate through nucleus before they can be experimentally detected. Thus experimentalists would rather like to speak about QE-like events defined by the condition that there are no mesons in the final state or SPP-like events with only one pion and no other mesons in the final state. It is clear that there is a significant difference between QE and QE-like events because the latter include  those in which pions were produced in the initial interaction but were later on absorbed inside the nucleus. It is also possible that rescattering of a nucleon from QE primary interaction results in pions in the final state. \\
\\
In direct experimental analysis further complication comes from the detection thresholds for kinetic energies of various species of particles. These can be very different. In the water Cherenkov detectors like SuperKamiokande they are $\sim 75$~MeV for $\pi^\pm$ and $\sim 485$~MeV for protons while in the ND280 near detector in the T2K experiment they may be about $100$~MeV and $150$~MeV correspondingly. Thus the measurement of QE-like and SPP-like cross-sections  require also some information from MC events generators and the uncertainty introduced by them vary from experiment to experiment. FSI effects are an important source of uncertainty of MC events generators and thus in the data analysis as well. This is why recently the experimental groups tend to provide the neutrino-nucleus cross-section data (with the FSI effects included) rather than neutrino-nucleon ones. 
Such data include corrections for detector efficiency and the effort is done to make them independent (as much as possible) from the nuclear physics assumptions of the MC codes used in the analysis. \\
\\
Many nuclear physics complications are absent in experiments performed on the hydrogen or deuterium target. This is the reason why there is still an interest in old bubble chamber experiments like ANL or BNL and surprisingly they still can be a source of interesting information \cite{deuterium}.  The advantage of new experiments like MiniBooNE and SciBooNE is in much higher statistics and better understanding of the neutrino flux. \\
\\
In my presentation I will include many  comments on the status of Monte Carlo events generators. It is important that the generators are modernized in parallel with better understanding of $\sim 1$~GeV neutrino interactions. 

\section{Quasielastic axial mass}

Charge current quasielastic reaction (CCQE) is the dominant process in the sub-GeV neutrino energy region typical for MiniBooNE, SciBooNE or T2K experiments. The theoretical description of neutrino-nucleon reaction is based on the conserved vector current (CVC) and the partially conserved axial current (PCAC) hypotheses. As a result of simple analysis the unique unknown quantity is the axial form-factor $G_A(Q^2)$ for which one typically asssumes the dipole form with only one unknown parameter called axial mass $M_A$. If the deviations from the dipole form of $G_A$ are of similar size to those in the case of electromagnetic form-factors it would be very difficult to detect them and the basic assumptions described above seem to be well justified. Thus the aim of CCQE experiments is to measure the value of $M_A$. Even if in the experiments neutrinos interact with bound nucleons the reported results should always refer to the parameter in the formula for free nucleon scattering. Obviously, any such measurement done on a nucleus target contains a bias from the model of nucleus used in the data analysis.\\
\\
The measurements of $M_A$ typically focus on the analysis of the shape of the differential cross-section in $Q^2$ that turns out to be sensitive enough for quite precise evaluation of $M_A$. The investigation of only the shape of the distribution of events in $Q^2$ has an advantage that it does not rely on the knowledge of the overall neutrino flux that usually carrries much uncertainty. The dependence of the total cross-section on $M_A$ can also be used as a tool to fix its value. The limiting value of the CCQE cross-section as $E_\nu\rightarrow\infty$ can be calculated in the analytical way assuming dipole vector and axial form-factors \cite{ankowski}. In the exact formula the dependence on $M_A$ is strictly speaking quadratic but in the physically relevant region it is with a good approximation linear. It is an important fact that if the value of $M_A$ is increased e.g. from $1.03$ to $1.23$~GeV the cross-section and the expected number of CCQE events is raised by $\sim 20\%$.\\
\\
In the past there were several measurements of $M_A$  mostly on the deuterium target and until few years ago it seemed that the results converge to a value of the order of $1.03$~GeV. There is an additional argument in favor of a similar value of $M_A$ coming from the weak pion-production at low $Q^2$. PCAC based evaluation gives the value of $1.077\pm 0.039$~GeV \cite{MA_PCAC}. On the contrary all (with the exception of the NOMAD experiment) more recent high statistics measurements of $M_A$ report much larger values: K2K (oxygen, $Q^2>0.2$~GeV$^2$) $\rightarrow 1.2\pm 0.12$ \cite{k2k_oxygen_MA}; K2K (carbon, $Q^2>0.2$~GeV$^2$) $\rightarrow 1.14\pm 0.11$\cite{k2k_carbon_MA}; MINOS (iron, $Q^2>0$~GeV$^2$) $\rightarrow 1.19\pm 0.17$; MINOS (iron, $Q^2>0.3$~GeV$^2$) $\rightarrow 1.26\pm 0.17$\cite{minos_MA}; MiniBooNE (carbon, $Q^2>0$~GeV$^2$) $\rightarrow 1.35\pm 0.17$; MiniBooNE (carbon, $Q^2>0.25$~GeV$^2$) $\rightarrow 1.27\pm 0.14$ \cite{MB_MA} (for completness: NOMAD (carbon, $Q^2>0$~GeV$^2$) $\rightarrow 1.07\pm 0.07$ \cite{nomad_MA}).\\
\\
There are a few possible explanations of the discrepancy. In the simplest, one notices large uncertainties of the measured value of $M_A$ and treats the discrepancy as merely statistical fluctuations (after all the effect is on the $< 2\sigma$ level). The problem is that there are several independent measurements. In the case of MiniBooNE large values of $M_A$ were obtained from the investigation of the shape of the distribution of events in $Q^2$ and also as a fit to the normalized cross-section and both evaluations do agree. This weaken doubts that are sometimes raised concerning the MiniBooNE's understanding of the overall normalization (integrated flux). In fact, there are other MiniBooNE measurements e.g. (CC1$\pi^+$) giving rise to larger than expected cross-section. The normalization (flux and fully correlated systematic errors) uncertainty is evaluated by the MiniBooNE collaboration to be $10.7\%$. A delicate element of the MiniBooNE's data analysis is a subtraction from the sample of QE-like events of those that are believed to be not QE in the primary interaction. In the analysis the NUANCE \cite{nuance} MC event generator based on the Fermi gas model was used. Obviously such subtraction depends on assumptions of the MC model. MiniBooNE collaboration {\it corrected} the MC prediction for this background by the function that was obtained by comparing a sample of SPP-like events to the predictions of the same MC generators. The shape of the correction function is rather poorly understood but it has an obvious and quite important impact on the extracted value of $M_A$. The function quantifies a lack of precision in describing processes like pion absorption and this can have different effect on understanding of QE-like and SPP-like samples of events.\\
\\
A separate difficulty is related to the low $Q^2$ region. It has been known for many years that MC events generators have problems with correct reproduction of the shape of differential cross-section in this region ({\it low} $Q^2$ means typically $Q^2<\sim 0.1$~GeV$^2$). This is the reason why in the data analysis very often (see above) some cuts are imposed. The low $Q^2$ problem has to do with the validity of the impulse approximation. We know from the electron scattering that for momentum transfers $q\leq 350-400$~MeV/c the models based on IA fail to agree with the data. In this region collective nuclear effects become important and computational techniques like RPA or CRPA should be used \cite{rpa_luis}. Since $q>\omega$ where $\omega$ is the energy transfer, and $Q^2=q^2-\omega^2$, the region of the failure of IA is clearly contained in the domain  $Q^2< 0.1$~GeV$^2$. How large is this {\it dangerous} region? Contrary to what might be expected, in the case of neutrino interactions it is always large, of the order of $15\%-20\%$ of the total cross-section, independently on the neutrino energy (for energies $E_\nu<500$~MeV it is even larger) \cite{artur_ia}. Experimental groups invented some ad hoc solutions to deal with the low $Q^2$ problem. The MiniBooNE collaboration proposed an effective  parameter $\kappa$ to increase the effect of Pauli blocking \cite{kappa}. CCQE fits were done simultaneously to $M_A$ and $\kappa$, treated as free parameters. In the recent MiniBooNE's paper \cite{MB_MA} the best fit for $\kappa$ is within $1~\sigma$ consistent with $\kappa =1$ (means no modification of the Pauli blocking). It is important that the 1-parameter fits for $M_A$ (with $\kappa =1$) do not lead to significantly different results. Also the MINOS collaboration proposed an ad hoc modification of the Pauli blocking \cite{minos_MA}.\\
\\
An interesting theoretical idea to explain the $M_A$ value discrepancy comes from the sophisticated many-body nuclear model proposed about 10 years ago by J. Marteau \cite{marteau} and developed recently by M. Martini \cite{martini}. This is the non-relativistic model that includes QE and $\Delta$ production primary interactions, RPA corrections and  local density effects. The most interesting feature of the model is the evaluation and inclusion of elementary 2p-2h excitations. This contribution goes beyond the IA and is absent in free nucleon neutrino reaction. There is the strong evidence from electron scattering data (even off light nuclei like $^4He$) that in the transverse response function the contributions from one and two body mechanisms are similar in strength \cite{Carlson}. The 2p-2h contribution is quite large and it is claimed to be able to account for the large CCQE cross-section as measured by the MiniBooNE collaboration, In the case of neutrino-carbon CCQE process after averaging over the MiniBooNE beam, nuclear effects are expected to increase the cross-section from 7.46 to 9.13 (in the units of $10^{-39}$cm$^2$). This includes a reduction of the cross-section due to RPA effects and increase thanks to 2p-2h contribution. In the case of antineutrino-carbon CCQE reaction the RPA and 2p-2h effects cancell each other approximately, and the cross-section is virtully unchanged (modification from 2.09 to 2.07). We see that it will be very important to compare predictions of the model with anti-neutrino CCQE MiniBooNE data. It will take about 1-2 years before MiniBooNE CCQE anti-neutrino data analysis is completed \cite{mb_anti}. It is also  important to compare predictions from the model with MiniBooNE's CCQE double differential cross-section data for the distribution of events in muon kinetic energy and scattering angle \cite{MB_MA}. 2p-2h contribution is parametrized in terms of energy and momentum transfer (there are 2 different parametrizations) which can be translated into double differential cross-section. The hope is that when the extra contribution is superimposed on the IA QE events with {\it standard} $M_A\sim 1.05$~GeV the overall distribution mimics the pure QE one with $M_A\sim 1.25$~GeV.\\

\section{Neutral current $\pi^0$ production}

Neutral current $\pi^0$ production (NC$\pi^0$) is a dangerous background to $\nu_e$ appearance oscillation signal in Cherenkov detectors and during the last $\sim 5$ years there were several attempts to measure its cross-section. Since we are interested in $\pi^0$ leaving nucleus, the experimental data is always given in this format with all the FSI effects included. If this definition of NC$\pi^0$ is adopted the relevant events can origin from: NC$1\pi^0$ primary interaction with $\pi^0$ not affected by FSI, NC$1\pi^+$ primary interaction with $\pi^+$ being transformed into $\pi^0$ in charge exchange FSI reaction, ... It is clear that a comparison of theoretical models with such results is rather difficult. On the other hand, such format of the data is very useful to check the performance of MC generators of neutrino interactions and tools like GiBUU \cite{gibuu}. An additional challenge related with the NC$\pi^0$ production is that it includes a COH contribution, which in the case of MiniBooNE's beam neutrino-carbon reactions ($<E_\nu> \sim 1.2$~GeV)  is estimated to be on the level of $\sim~20$\% \cite{mb_coh}.\\
\\
Four recent measurements of NC$\pi^0$ (K2K \cite{k2k_ncpi0}, MiniBooNE neutrinos, MiniBooNE antineutrinos \cite{mb_ncpi0} , SciBooNE \cite{sb_pi0}) are complementary. They use three different beams (K2K, FermiLab Booster neutrino and anti-neutrino beams) and the targets: $H_2O$ (K2K), $CH_2$ (MiniBooNE) and $C_8H_8$ (SciBooNE). MiniBooNE presented the results in the form of absolutely normalized cross-section while K2K and SciBooNE reported only the ratio $d\sigma (NC1\pi^0)/ \sigma (CC)$. There is an important difference in the measured quantity: K2K and MiniBooNE present their results as measurements of the final states with only one $\pi^0$ and no other mesons. SciBooNE speaks about states with at least one $\pi^0$ in the final state i.e. a contamination from $1\pi^01\pi^\pm$, $2\pi^0$ and $>2\pi$ (with $>1\pi^0$) final states is included (evaluated with NuWro to be on the level of 17\%). All the experimental groups present the final results in the form of flux averaged distributions of events in the $\pi^0$ momentum, and in the case of MiniBooNE and SciBooNE  in the $\pi^0$ production angle. Additionally, MiniBooNE provides also the data for the cross-section before subtraction of the contribution from the neutrino or anti-neutrino beam contaminations. This data represent the measurement that is in the maximal possible degree independent on the assumptions of the MC events generator.\\
\\
The comparison with the NC$\pi^0$ data can only be done within MC events generators. There are several ingredients of the MC that are tested simultaneously: NC $\Delta$ (or more generally: resonance) production, nonresonant contribution to SPP, modification of the $\Delta$ width in the nuclear matter, angular distribution of the $\pi^0$ from a $\Delta$ decay, a COH component, a multipion production, FSI effects (the $\pi^0$ absorption rate in the nucleus, cross-sections for pion charge exchange reaction in the nucleus, formation zone effects). It is an unfortunate situation that the separate ingredients of the MC are known with unsatisfactory precision.\\
\\
There is only one reported measurement of the NC$1\pi^0$ cross-section in the Gargamelle bubble chamber (the target was composed of $C_3H_8$ (90\%) and $CF_3Br$ (10\%)) with the $\sim 2$~GeV beam of neutrinos \cite{krenz} (we notice that the data from the paper \cite{krenz} contain also a non-negligible contribution from the COH reaction what makes the extraction of the resonance NC$\pi^0$ production even more complicated). The results were published in the form of efficiency corrected numbers of events in several exclusive SPP channels and also as nuclear effects corrected numbers of events on free nucleon targets. The way of treatment of the nuclear effects for the topologies of final states is described in \cite{anp}.  Later reanalysis of the data introduced absolute normalization (it was possible because in the original paper one CC SPP channel was  included in the analysis) and flux averaged cross-sections were obtained \cite{hawker}. An interesting feature of the final results is that the proton's NC$\pi^0$ cross-section is much larger than the neutron's one. Many details of the analysis are given in the original paper and it seems possible to reexamine nuclear effects using the better knowledge of pion absorption rates in various nuclei. \\
\\
The NC$\pi^0$ differential cross-section for pion production angle is very sensitive to the COH component that contributes to the forward directions only. The differential cross section for pion kinetic energy is very sensitive to FSI effects and in particular to the absorption rate dependence on pion momentum. A useful comparison of performance of FSI models in various MCs is done in \cite{ladek_project}. Authors of intranuclear cascade codes confront the models with the available $\pi^+$ nucleus scattering data \cite{dytman}. There is quite a lot of $\pi^+- {}^{12}C$ reaction (with the separate absorption and charge exchange rates) data in the $\pi^+$ kinetic energy range $50-500$~MeV and the typical uncertainty is $\sim 20$\% \cite{pion_nucleus}. The data for $\pi^+- {}^{16}O$ absorption is scarse and it is not obvious how the reaction cross-sections scales as the size of a nucleus increases. Other observables that are useful in benchmarking FSI models are nucleon's \cite{nucl_trans} and pion's   \cite{pion_trans} \cite{cosyn} transparencies.

\section{Coherent pion production}

In the coherent pion production (COH) the target nucleus remains in the ground state. Four possible channels are possible, for CC and NC reactions, for neutrinos and for anti-neutrinos, see Eq.  (\ref{coh}). The NC COH cross-section is believed to contribute significantly to the overall NC$\pi^0$ cross-section and the process has been a subject of intensive experimental and theoretical investigation. There is a clear experimental signal for the COH reaction at higher energies and the aim of recent measurements is to fill a gap of the knowledge of about $\sim 1~GeV$ COH cross-section. In the case of CC reaction K2K \cite{k2k_coh} and SciBooNE \cite{sb_coh_cc} reported no evidence for the COH component. In the case of NC reaction MiniBooNE \cite{mb_coh} and SciBooNE \cite{sb_coh_nc} detected the COH component. The identification of the COH signal is done based on the predictions of the MC generatotor of events. In the case of CC COH reaction the sample of $1\pi^+$ events was searched at low $Q^2$, and for the NC reaction the distributions of observed $\pi^0$ in the forward direction was analysed. \\
\\
It is interesting to look at the numerical values obtained in the measurements. For the K2K's analysis of the CC COH reaction, the 90\% confidence limit for the upper bounds of the COH cross-sections on carbon was estimated to be $0.6\%$ of the total CC cross-section.
Similarly, the SciBooNE's results (also for the carbon) are: $0.67\%$ at $<E_\nu>\sim 1.1$~GeV, and $1.36\%$ at $<E_\nu>\sim 2.2$~GeV.
For the NC reaction, thanks to the information about recoil protons, SciBooNE \cite{sb_coh_nc} evaluated the ratio of the COH NC$\pi^0$ production to the total CC cross-section as $(1.16\pm 0.24)\%$. 
SciBooNE reported the measurement also in the form of the ratio of CC COH $\pi^+$ to NC COH $\pi^0$ production, estimated as $0.14^{+0.30}_{-0.28}$. The value is surprisingly low and is a challenge for theoretical models. For the NC reaction MiniBooNE also evaluated the COH component (plus possible hydrogen diffractive contribution about which little is known) in the NC$\pi^0$ production as 19.5\% (at $<E_\nu>\sim 1.2$~GeV) and then the overall flux averaged NC$1\pi^0$ cross-section as $(4.76\pm0.05\pm0.76)\cdot 10^{-40}$cm$^2$/nucleon. Unfortunately, it is difficult to translate both measurements into the absolutely normalized value of the NC COH cross-section. The reason is that the first from the above two measurements strongly depends on the details how is NC$\pi^0$ production modelled in the MC used by the MiniBooNE collaboration i.e. NUANCE \cite{nuance}. In NUANCE, RES, COH and BGR (background) NC$\pi^0$ reactions are defined according to the primary interaction. A peculiar feature of NUANCE is that COH produced pions are subject to FSI, in a similar way as pions produced in the RES primary interaction. In the final step of the analysis the fit is done for the composition of the sample of NC$\pi^0$ events in terms of three components, and the COH fraction is defined as $x_{COH}/(x_{COH}+x_{RES})$ \cite{sam}.\\
\\
NUANCE contains an independent diffractive (DIFFR) component of SPP coming from reactions on free nucleons \cite{diffr}. The experimental signal of pions produced by the DIFFR mechanism is similar to the COH one, and in the process of selection of COH events they are likely to be put to the same category. In \cite{diffr} the evidence is shown for the presence of DIFFR component in the high energy neutrino reactions, in the region of invariant hadronic mass $W>2$~GeV. For $W<2$~GeV the DIFFR component is expected to contribute to the nonresonant background and in the MC implementation it is important to avoid a danger of double counting. MiniBooNE estimated the DIFFR contribution as 16\% of the overall `coherent' cross-section but its impact on the reported value of the COH fraction is not very important.\\
\\
The evaluation of the COH contribution to the cross-section is always made by confronting the measurements with the predictions of MC events generators. Before the last NuInt09 workshop S. Boyd and S. Dytman initiated the theory and MC comparison project \cite{steve_nuint09}. In the case of COH reaction ten different computations were compared, four from MC generators (GENIE \cite{genie}, NEUT \cite{neut}, NUANCE, NUWRO \cite{nuwro}) and six from theoretical models. All the MC generators implemented the same theoretical model of Rein\&Sehgal \cite{rs_coh}. It was surprising that the predictions from the MCs differ by more than 100\% . For the CC COH reaction on carbon at $E_\nu=1.1$~GeV the predicted cross-sections (per nucleon in the units of $10^{-40}$cm$^2$) were: 2.33 (NEUT), 1.88 (NUANCE), 1.31 (NUWRO) and 0.85 (GENIE). Predictions from theoretical models were usually about 0.5, slightly larger were results from the computations of models of Alvarez-Ruso et al \cite{luis_coh} (0.8) and Martini et al \cite{martini} (0.66). \\
\\
It is sometimes believed that it is sufficient to correct the implementation of the  model \cite{rs_coh} by a correction factor (e.g. 2/3 as it was done by MiniBooNE in the case of NUANCE). But the comparison of differential cross-sections in pion kinetic energy revealed even more dramatic differences between MCs \cite{steve_nuint09}. Theoretical computations predict smooth distributions with a maximum at $T_{\pi^0}\sim 150$~MeV (the actual position varies from $110$~MeV to $180$~MeV). The shape is correctly reproduced by NUANCE, while GENIE, NEUT and NuWro predict a lot of extra structure: a local minimum at $T_{\pi^0}\sim 170~MeV$ (presumably from pion absorption, GENIE, NuWro) or at $T_{\pi^0}\sim 300~MeV$ (NEUT). The large differences between MCs and theoretical models are clearly seen also in double differential cross-section at fixed pion production angle. \\
\\
Higher neutrino energy ($E_\nu >\sim 2$~GeV) COH production  (including recent NOMAD measurement) was successfully explained with the help of the PCAC based model \cite{rs_coh}. Adler's theorem relates $\sigma_{COH}(\nu + X\rightarrow \nu + X + \pi^0)$ at $Q^2\rightarrow 0$ with $\sigma(\pi^0 + X \rightarrow \pi^0 + X)$. Recently the model for the CC reaction, has been refined \cite{coh_pcac_mass} to include charged lepton mass effects. The new model predicts the $\sigma_{COH} (\pi^+)$/$\sigma_{COH} (\pi^0)$ ratio at $E_\nu=1$~GeV to be $1.45$ rather than $2$. Another very important modification was to use the available data for $d\sigma(\pi +\ {}^{12}C \rightarrow \pi +\ {}^{12}C$/$dt$ in the region of pion kinematical energy $100$~MeV$<T_\pi<900$~MeV.  As a result the predicted COH cross-section becomes smaller by a factor of 2-3. The other PCAC based approach is proposed in \cite{coh_pcac_pion}. The microscopic models for the COH reaction \cite{coh_micro} \cite{luis_coh} \cite{martini} assume $\Delta$ dominance and are expected to be more reliable at low neutrino energy. Within microscopic models there are still various approaches e.g due to differences in the treatment of a nonresonant background. The absolute normalization of the predicted cross-section depends on the adopted value of the $N\rightarrow \Delta$ form factor $C^A_5(0)$.

\section{Other measurements}

For completness I list and comment some other recent neutrino cross-section measurements and phenomenological investigations.\\
\\
MiniBooNE collaboration measured the ratio $\sigma (CC1\pi^+)/\sigma (CCQE)$ \cite{mb_ratio} which is an important observable,  independent on the overall neutrino flux normalization uncertainty. In the MC independent version the reported observable is the ratio of SPP-like to QE-like cross-sections as a function of neutrino energy. The uncertainty of this measurement is still larger than $10$\%, and the reason is the error in the recontructed neutrino energy. Using information contained in MC model it was also possible to evaluate the ratio of SPP to QE defined as primary interactions on an abstract isoscalar nuclear target. From the MiniBooNE data it follows that in the Monte Carlo events generators an increase of the axial mass $M_A$, which controls the value of the CCQE cross-section,  must be accompanied by a change of parameters that determine the value of CC$1\pi^+$ production.\\
\\
MiniBooNE reported preliminary results for the CC$1\pi^+$-like cross-section \cite{mb_piplus}. Thanks to the possibility to detect final $\pi^+$, it was possible to reconstruct neutrino energy and invariant hadronic mass $W$, based on muon and pion observables only. $W$ distribution shows a clear $\Delta$ peak. Both the total and differential cross-sections disagree with the MC (NUANCE) predictions and the data can be very useful for the future improvements in the numerical codes.\\
\\
Almost all the MC events generators rely on the Rein-Sehgal model for pion resonance (RES) production \cite{rs_res}. The model includes contributions from 18 resonances and covers the region $W<2$~GeV. In the original model, the charged lepton is assumed to be massless. The RS model should be modified in order to be more reliable in the sub-GeV neutrino experiments. Lepton mass corrections can be introduced following the prescriptions proposed in \cite{rs_lepton}. There are also arguments in favour of modifying form-factors used in the model \cite{rs_ff}. It is known that the vector part does not reproduce the electron scattering data and consequently also the axial part should be fixed by making a fit to the deuteron and/or MiniBooNE pion production data. In the paper \cite{jarek} a comparison study of some theoretical models \cite{rs_ff} \cite{olga} \cite{hernandez} with the MiniBooNE data was done. \\
\\
MiniBooNE reported also the results with NC elastic reaction cross-section \cite{mb_elastic}. The measurement was possible because the MiniBooNE detector, which is basically a Cherenkov detector, can observe also scintillation light from low momentum nucleons. An attempt is done to measure the strange quark component of nucleon spin using the proton enriched sample of events.

\section{Monte Carlo generators}

The performance of MC events generators cannot be better than the precision of neutrino cross-section measurements and this is on the level of 15\%-20\%. This explains the differences in the performance of MCs \cite{ladek_project}. Most important MCs (GENIE \cite{genie}, NEUT \cite{neut}, NUANCE \cite{nuance}, FLUKA \cite{fluka}) are principally aimed to be tools in the data analysis and their developments are guided by the needs of particular experiments \cite{gallagher}. Each of them has some priorities determined by the physical program and by the detection techniques. From the experimental point of view it is useful that the performance of a MC can be fine-tuned to a particular experimental situation and that one can rescale contributions from some exclusive channels in order to get an agreement with what is observed. From time to time there are discussions (e.g during the first NuInt01, and also during the last NuInt09 \cite{mc_discussion}) about the construction of a universal Monte Carlo generator of neutrino interactions. Few years ago a project aimed in this direction was launched under the name of GENIE ({\it Generates Events for Neutrino Interaction Experiments}) \cite{genie}. It is written in C++ and its architecture allows to absorb various pieces of codes from other MCs. In the core of GENIE there is another MC generator NEUGEN \cite{neugen}, and there is still an intensive work on various elements of the code. \\
\\
It seems however that in the perspective of 5 years several MCs will be developed in parallel,  what is probably useful for their quality: it will be possible to compare and cross check the performance of proposed improvements. In the past many MC comparison tests were done and presented during NuInt workshops \cite{nuint}. Currently, in the T2K collaboration, several useful GENIE-NEUT  comparison studies are performed. It is also important that there are MCs or MC-like tools developed by theorists (GIBUU \cite{gibuu}, NuWro \cite{nuwro}) which allow for more flexible manipulations with various ingredients of the codes. MCs are large and long-term projects and it is often difficult to verify quickly all the consequences of even minor modifications in order to avoid risks of spoiling the self-consistency of the code.\\
\\
There are very few {\it obvious} modifications which definitely should lead to an improvement of the performance of MCc. One of them is, as it was argued earlier, an implementation of a better model for the COH pion production. Another one seem to be the replacement of the Fermi gas (FG) model by a more reliable formalism in the description of a nucleus target. \\
\\
Fermi gas model is determined by two parameters: Fermi momentum and binding energy and it defines the probability distribution of finding inside nucleus a nucleon with a given value of momentum (quadratic distribution) and binding energy (a constant value). From the electron scattering experiments it is known that in the interesting kinematical region the FG model allows for reasonable agreement with the data. The advantage of the model lies in the simplicity. Also its MC implementation is straghtforward. Is is however  known from the more detail analysis of the electron scattering data that the FG model suffers from many limitations and is unable to reproduce separately longitudinal and transverse nuclear response functions.\\
\\
There is a variety of approaches to describe nuclear effects in electron scattering which were later applied to neutrino interactions. Short descriptions of many of them is given in \cite{steve_nuint09} \cite{luis_review} together with a list of basic references. Monte Carlo implementation of most of the theoretical models is usually  complicated, if not impossible. MC requires a propagation of all the degrees of freedom (particles taking part in interactions) and it is not sufficient to know the inclusive neutrino interaction cross-section. There is however a model that can replace FG and which allows for much better agreement with the data. It is called {\it spectral function} (SF) \cite{spectral_function} but strictly speaking the name {\it hole's spectral function} should be used. The use of SF has been advocated by Omar Benhar during several NuInt workshops. SF is defined as a joint probability distribution to find inside nucleus a nucleon with a given momentum and binding energy. SF arises naturally as one calculates the neutrino quasielastic cross-section in the plane wave impulse approximation (PWIA) \cite{pwia} i.e. assuming that the nucleon arising in the final state after primary interaction leaves the nucleus with no FSI effects. Together with a model for FSI \cite{sf_fsi} (or for the particle spectral function) the SF model leads to a very good agreement with the electron-nucleus cross-section data for momentum transfers larger than $\sim 350$~GeV \cite{sf_electron_data}. The available models of SF combine information from the mean field theory (shell model) and a contribution from short range correlations (SRC). The shell model structure is clearly seen as the peaks in the probability distribution corresponding to shell model orbitals. SRC part contributes to a large nucleon momentum tail in the probability distributions. SF for few important nuclei (carbon, oxygen, iron) were constructed by Omar Benhar and his collaborators. There also exist  approximate models of SF for medium size nuclei like calcium and argon, which were shown to lead to a good agreement with the electron scattering data \cite{ankowski_sobczyk_2}. It is estimated  that correlated proton-neutron pairs are $\sim 18$ times more likely than proton-proton ones, and that $\sim 25\%$  of nucleons inside medium size nuclei have momenta larger than $300$~MeV \cite{src}. Neutrino QE interactions may provide an opportunity to observe correlated pairs because the signal (up to distortions caused by FSI effects) would be a pair of high momentum protons possible to observe e.g. in liquid argon detectors. \\
\\
The implementation of SF formalism into MC events generators is not trivial. SF introduces extra integrations which, if not performed in the numerically efficient way, can make the process of generating events very time consuming. For that reason a few years ago an {\it effective} approach was proposed \cite{ankowski_sobczyk_1}. The genuine SF defines the momentum distribution and  also the momentum dependent binding energy, which is defined so that it replaces the constant binding energy of the FG model. It was shown that the effective approach provides a good approximation of the SF \cite{ankowski_sobczyk_1}. Its implementation is easy unless the MC code (that is the case for NEUT) uses analytical formulas obtained by averaging elements of the hadronic tensor over target nucleon momentum within the FG model. Recently, the SF formalism was implemented in the MC generator NuWro \cite{juszczak_sobczyk}. It has been already known that SF does not introduce much change, with respect to FG, as far as the shape of the QE differential cross-section in $Q^2$ is concerned \cite{ankowski_benhar_farina}. In \cite{juszczak_sobczyk} it is shown that the fit to the value of the axial mass $M_A$ based on the MiniBooNE's double differential cross-section in muon kinetic energy and scattering angle done within the SF model leads to a very similar value as the FG model ($M_A\sim 1.4$~GeV). Similar results were obtained also by the MiniBooNE collaboration \cite{laird}. In the case of FG model the goodness of fit is better because the use of SF reduces  the predicted value of overall CCQE cross-section by about $20\%$. The elimination of bins with large contribution of events with small values of momentum transfer (more than 50\% of events with $|\vec q|<350$~MeV) makes the fitted value of $M_A$ smaller by about $100$~MeV. The modification goes in the desired direction but there remains a large gap with respect to the old deuterium bubble chamber measurements.

\section*{Acknowledgements}
I would like to thank Agnieszka Zalewska and other organizers of the Cracow Epiphany Conference for the kind invitation to give this talk.\\
\\
The author was supported by the grant 35/N-T2K/2007/0 (the project number DWM/57/T2K/2007).

\end{document}